# A New Model of Chemical Bonding in Ionic Melts

Vitaly V. Chaban[1], Yuriy V. Pereverzev, and Oleg V. Prezhdo

University of Rochester, Rochester, New York 16427, United States

**Abstract**. We developed a new physical model to predict macroscopic properties of inorganic molten systems using a realistic description of inter-atomic interactions. Unlike the conventional approach, which tends to overestimate viscosity by several times, our systems consist of a set of ions with an admixture of neutral atoms. The neutral atom subsystem is a consequence of the covalent/ionic state reduction, occurring in the liquid phase. Comparison of the calculated macroscopic properties (shear viscosity and self-diffusion constants) with the experiment demonstrates good performance of our model. The presented approach is inspired by a significant degree of covalent interaction between the alkali and chlorine atoms, predicted by the coupled cluster theory.

[1] Corresponding author. E-mail: v.chaban@rochester.edu, vvchaban@gmail.com

**Introduction**

Ionic compounds (ICs)[1-6] play an important role in modern chemical industry and the fundamental studies of atomistic interactions[5, 7-10]. Due to the very strong attraction between the particles (more than 200 kJ/mol)[11], ICs exist in the solid state at room temperatures, and their melting points are normally above 1000 K. Room-temperature ionic liquids present an important exception. They exist in the liquid state because of the bulky organic cations with a greatly delocalized positive electrostatic charge[12, 13].

It is widely believed that inter-ionic interactions in most inorganic ICs, especially those composed of alkali and halogen atoms, are purely ionic. Thus, the potential energy of the system can be calculated simply as

$$U = \sum_{i,j} \frac{q_i^+ q_j^-}{r_{ij}} + \sum_{i,j} \frac{q_i^+ q_j^+}{r_{ij}} + \sum_{i,j} \frac{q_i^- q_j^-}{r_{ij}} + \sum_{i,j} \frac{C_{ij}}{r_{ij}^{12}},\qquad(1)$$

where | q | can equal to 1, 2, and rarely, 3 elementary charge units, $C_{ij}$ is the constant of electron-electron repulsion between any pair of atoms, and *i*, *j* are atom indices. This relation is applicable to all condensed phases of the substance and plays an important role in the computations of thermodynamics properties, phase transition points, and other types of molecular simulations[5, 7, 14]. However, the concept of a purely ionic bond does not completely agree with quantum mechanics, as applied to these many-electron systems.

In the present work, we report a significant degree of covalence in the bonded pairs of NaCl and KCl at zero (ideal geometry) and finite temperatures. Consequently, the common model of chemical bonding in sodium and potassium chlorides is revised. Upon melting, a reduction of the covalent/ionic state takes place, engendering both ionic and neutral subsystems. The concentration of the latter one is determined by the degree of covalence in the covalent/ionic state. The comparison of our results *versus* experimental data on shear viscosity of NaCl and KCl melts strongly supports this development. We start from the high-level post-Hartree-Fock computations of the optimized ion pairs in vacuum, in order to assess quantitatively the degree of covalence. The obtained data are further utilized to develop a new interaction model. Finally, classical molecular dynamics (MD) simulations of molten salts are carried out, providing shear viscosities and ionic self-diffusion constants at 1080 K.

**Details of Numerical Simulations**

The research starts with an *ab initio* analysis of the degree of covalence in the bonded Na+Cl and K+Cl pairs as a function of inter-atomic distance. These data lay the groundwork of the new model of atomistic interactions, in which neutral particles dynamically replace a certain fraction of charged particles.

The interaction energies between the ions and the ion partial charges (Figures 1-2) are based on the wave-functions obtained by the coupled cluster theory with double substitutions (CCD) from the Hartree-Fock determinant. The comprehensive aug-cc-pVTZ and 6-311++G** basis sets are used for sodium and potassium chlorides, respectively. The bond energies of the atomic pairs (energies of interaction) are calculated according to the following formula, $E_{int} = E_{XY} - E_{X+} - E_{Y-}$, where $E_{X+}$ is the total energy of the cation in vacuum, $E_{Y-}$ is the total energy of the anion in vacuum, and $E_{XY}$ is the total energy of the corresponding ion pair. The obtained interaction energy is further divided into components, assuming that it consists of (1) pure electrostatic interaction, (2) van der Waals interaction, and (3) covalent interaction. The electrostatic fraction ($E_{coul}$) is found *via* the Coulomb law using CHELPG partial charges[15] as described below. The van der Waals energy ($E_{VDW}$) is based on the Lennard-Jones (12, 6) parameters in the OPLS *ab initio* based force field (FF)[16]. Finally, the covalent interaction is calculated as $E_{cov} = E_{int} - E_{VDW} - E_{coul}$.

The partial atomic charges are assigned using the Hirshfeld procedure (based on the electronic density distribution) and electrostatic potential fit (CHELPG scheme), the latter constrained to rigorously reproduce the *ab initio* dipole moment of the system. Additionally, the Mulliken population analysis is presented for qualitative comparison. Since the Mulliken approach is based on the equal division of the off-diagonal terms between the two basis functions, leading to very rough results, we base our quantitative conclusions only on the Hirshfeld and CHELPG partial charges.

The dynamical properties of the salts melts (Figures 3-4) are derived using atomistic-precision dynamics simulations, where inter-atomic interactions are described by the pairwise Coulomb and Lennard-Jones (12, 6) terms. The electrostatic charges on the cations and anions are +1 and -1, respectively, whereas the sigma and epsilon parameters for the Lennard-Jones equation are borrowed from the OPLS force field[16]. The properties of all NaCl and KCl melts are obtained from 5,000 ps long trajectories, generated using the leap-frog algorithm with the 0.001 ps time-step. The atomic coordinates and energy terms are saved every 0.01 ps for subsequent analysis. An appropriately modified GROMACS molecular dynamics engine[17-19] is used for these calculations. The experimental data for the NaCl and KCl melts at the specified temperatures are reported by Janz elsewhere.[20]

Each simulated system contains 1000 cations ($Na^+$ or $K^+$) and 1000 anions ($Cl^-$), where a certain fraction (Figures 3-4) of the ions is dynamically substituted by neutral atoms. We assign neutral atoms according to the following procedure. (1) A neighborhood criterion is defined as $r_{X+Y-} < r_{1st\ min.\ RDF(X+Y-)}$, where $r_{X+Y-}$ is distance between any cation and anion, and $r_{1st\ min.RDF(X+Y-)}$ is the position of the first minimum in the radial distribution function for these particles calculated in the "fully ionic" system. (2) After a certain time interval, all species in the system are scanned, and a table identifying neighboring particles is created. (3) A pseudorandom number generator is initialized in order to select a certain fraction of ion pairs to be substituted by neutral atoms. (4) Every time interval defined above, the neutral atoms are reassigned using the newly generated table of neighboring particles. In the present study, we apply the time interval of 25 fs, which is the timescale of the fastest nuclear motion available in these systems. We also probed the 50 fs and 100 fs time intervals, and observed no changes in the resulting dynamical properties. However, if the value is increased to 100 ps, neutral atoms tend to form clusters (non-polar islands), whose dynamics as a whole is certainly unphysical.

The diffusion constants are calculated *via* the well-known Einstein relation starting from the mean square displacements of each atom. The shear viscosities are derived *via* non-equilibrium simulations assuming that the energy deposited into the system by external forces is dissipated through viscous friction. The generated heat is removed by coupling to a thermostat. For this method to work properly, the shear rate should not be so high that the system gets too far from equilibrium[21]. Based on the densities of our systems and the expected viscosity values, we apply the acceleration of 0.05 nm ps$^{-2}$. Probing smaller acceleration values did not change the results.

**Derivation of the Model**

The energy of the Na-Cl and K-Cl ionic bonds and the partial atomic charges are depicted *versus* the ion separation distance in Figures 1-2. Using electron density provided by the coupled cluster theory, Hirshfeld partial charges are assigned to each atom, illustrating a 20-30% electron transfer (ET) to the cation at the energy minimum. In other terms, rather than forming a purely ionic bond, the atom pairs create mixed ionic/covalent bonds, with the degree of covalence ranging from ~30% in Na-Cl to 20% in K-Cl (20%). Interestingly, CCD predicts rather significant charge transfer (5%) even at the inter-atomic separation of 5Å. If the distance between the ions is increased to 7Å and more, no ET is seen. This result is very important for condensed phase simulations of ionic and ion-molecular systems. Mulliken analysis repeats the qualitative trend well, although it assigns somewhat smaller absolute charges, as should be expected. The interaction energy curves of NaCl and KCl (Figure 2a) are very similar in both shape and values. The bond dissociation energies of ~500 kJ/mol are comparable with those of common diatomic molecules with weakly polar and non-polar covalent bonds (400-1000 kJ/mol). It should be underlined that in the context of melts we use the total energies of ions, $E$ (Na$^+$) and $E$ (Cl$^-$), rather than of atoms, $E$ (Na$^0$) and $E$ (Cl$^0$), in order to estimate the interaction energies in the alkali-halide pairs. The difference between $E$ (Na$^+$) + $E$ (Cl$^-$) and $E$ (Na$^0$) + $E$ (Cl$^0$) is 152 kJ/mol, neutral atoms being certainly more stable. For estimation of the isolated pair bonding, the aforementioned correction energy has to be implemented.

In order to analyze the structure and nature of the investigated bonds, we decompose the total pairwise interaction energy into the Coulomb, covalent, and dispersion contributions (Figure 2b). The dispersion, calculated according to the *ab initio* based OPLS force field, is ~2 kJ/mol, *i.e.* it presents a negligible contribution compared to the other components. One can claim that classical molecular dynamics simulations of ionic melts can be successfully accomplished without the Lennard-Jones interaction term. This is different, for instance, from room-temperature ionic liquids, for which the dispersion contribution is noticeable[11]. To further differentiate between the covalent and ionic contributions to the energy, we employ the CHELPG *ab initio* electrostatic potential fitting scheme. Although all charge assignment schemes provide somewhat arbitrary partial charges, the difference between the best of them does not exceed a few percent and, therefore, it is not essential. At the interaction energy minimum, the contributions of the Coulomb interaction are 72% (NaCl) and 82% (KCl). As the separation increases, the electrostatic contribution grows up to 100%, *i.e.* covalent bond disappears. The lower fraction of covalent bonding in KCl compared to NaCl is probably due to its greater electronic polarizability dictated by the larger cation size.

Based on the aforementioned partial charge analysis, we can write down an expression for the valence electron state, $|e,\Psi\rangle$, in the ionic pair, for instance Na-Cl, as

$$|e,\Psi\rangle = A|e,Na\rangle + B|e,Cl\rangle, \qquad (2)$$

where $A$ and $B$ are amplitudes of the $|e,Na\rangle$ (electron occupies sodium atom) and $|e,Cl\rangle$ (electron occupies chlorine atom) states, respectively. The amplitudes are normalized assuming that $|A|^2 + |B^2| = 1$. The charges on the cation and anion in the $|e,\Psi\rangle$ state are equal to $\pm|Bq_e|^2$. Note that the structure of state $|e,\Psi\rangle$ (eq. 2) is valid rigorously for an isolated pair of atoms. In the crystalline state of salts, the distribution of the electron density, and consequently, the structure of covalence are different. Below, for simplicity we will apply eq. 2 to analyze the liquid state. Due to interaction with other particles in the molten salt, the bonding state of the pair, $|e,\Psi\rangle$, undergoes a collapse (or reduction) to either $|e,Na\rangle$ or $|e,Cl\rangle$ with the probabilities of $|A|^2$ or $|B|^2$. Thus, the electron density of the $|e,\Psi\rangle$ state shifts to either sodium (creating neutral atoms) or chlorine (creating ions). This transformation can be depicted by a free energy profile *versus* reaction coordinate (Figure 3). At low temperatures (crystalline state), a stable covalent/ionic state, $|e,\Psi\rangle$, with a single energy minimum exists. At higher temperatures, $|e,\Psi\rangle$ becomes metastable, engendering the states with neutral atoms and pure ions. One can suppose that under certain conditions the minimum at $|e,\Psi\rangle$ can be neglected.

The reported theoretical predictions are indirectly supported by the comparison between the experimental and simulated shear viscosities in molten NaCl and KCl salts. While the simulations based on the OPLS FF predict the viscosities of 3.9 and 4.0 cP for NaCl and KCl at 1080 K, the corresponding experimental values are 1.03 and 1.02 cP, respectively[20]. The factor of four difference cannot be attributed to the Lennard-Jones parameters of the ions. Therefore, the observed discrepancy is connected with the electrostatic potential of the liquid system. It is reasonable to assume that the simulated viscosity would decrease, if a certain fraction of charged particles, $Na^+$, $K^+$, $Cl^-$, is substituted by neutral atoms, Na, K, Cl, according to the aforementioned considerations.

Figures 4-5 show the dependence of shear viscosity upon the fraction of neutral atoms in the simulated systems at 1080 K. As predicted, even a small number of the dynamically assigned neutral atoms significantly lower viscosity, whereas a ~10 molar percent substitution produces excellent agreement between the experimental and computed values for both melts. The non-equilibrium concentrations of the charged and non-charged species ($C_n$ and $C_i$) evolve in time as

$$\frac{dC_n}{dt} = -k_{ni}C_n^2 + k_{in}C_i^2, \qquad (3)$$

where $k_{ni}$ is the rate of transition from neutral to charged particles, and $k_{in}$ is the rate of inverse transitions. In the equilibrium state, provided that $C_i + C_n = 1$, the coefficient K equals to

$$K = C_n^2 / (1-C_n)^2 = k_{in} / k_{ni}. \tag{4}$$

Eq. (3) can be used to estimate the equilibrium constant, once $C_n$ is available. The analysis presented above gives $C_n \approx 0.1$; hence, $K \approx 0.012$. Note that the equilibrium concentrations are directly connected to the amplitudes in $|e, \Psi\rangle$,

$$C_i \approx |B|^2, C_n \approx |A|^2. \tag{5}$$

According to the hole theory of liquid state developed by Eyring and co-workers[22, 23], momentum transfer within a liquid occurs due to transitions of particles between energetic states with a barrier $\Delta E$. The viscosity, $\eta$, equals to

$$\eta = F \exp[\Delta E / k_B T], \tag{6}$$

where $F$ is an empirical coefficient, $k_B$ is the Boltzmann constant, and $T$ is the absolute temperature. The activation energy is necessary to create a hole for a particle to enter the flow. The larger is the effective diameter of the particle, the higher is the energy required to create the hole. By significantly polarizing its environment, an ion requires a greater energy than a neutral atom, $\Delta E$ (eq. 6), to jump to a new position. Therefore, an increase in the concentration of neutral particles in the NaCl and KCl melts decreases the viscosity of the medium, as observed in the atomistic simulations (Figure 4). Note that it is not possible to differentiate between the diffusion constants of the neutral and ionic species, for example, $Na^0$ and $Na^+$, since their lifespan is much shorter than the time required to calculate diffusion constants based on atomistic trajectories. Thus, the diffusion constant of a certain species in a system can be expressed as the sum,

$$D = k_B T \left( \frac{C_i}{f_h + f_p} + \frac{C_n}{f_h} \right), \tag{7}$$

where $f_h$ is the conventional hydrodynamic friction and $f_p$ is the polarization friction, which influences the motion of charged particles exclusively.[24] As follows from eq. 7, a growth in the concentration of neutral particles should increase the diffusion constant (Figure 5).

One may be surprised that the best performing fractions of neutral particles (Figure 4) differ from the extent of the calculated electron transfer (Figure 1). There are at least three fundamental physical reasons for the difference. First, our *ab initio* calculations involve only a pair of atoms, whereas in the condensed phase the first coordination shell of a spherical ion includes about six counter-ions. It is not easy to predict the change in ET for this situation, since neighboring ions can both increase ET by further polarizing the electron density, and decrease it by shielding the interactions of counter-ions with one another. Periodic high-accuracy *ab initio* calculations of melts may be of a certain interest in this context. Second, the reported ET corresponds to the ideal geometry at zero temperature, certainly providing smaller inter-atomic distances than those present in the melts at 1080 K. Since larger separations lead to smaller ET (Figure 1), this consideration explains the observed discrepancy. Third, absent in the quantum mechanical calculations, temperature appears in the dynamics studies. Electronic motion is much faster than the motion of nuclei, and the ratio of the electronic and nuclear response time-scales should change with temperature. It is quite possible that moving nuclei lack enough time to respond to all changes in the electron density. Hence, the ET impact in the dynamics simulations at a finite temperature is smaller than that manifested by quantum mechanics at absolute zero. The interplay of these three factors determines the actual amount of ET in the IC melts, *i.e.* the coefficients in eq. 2. The ET can be estimated indirectly following our methodology and using the transport properties such as shear viscosity and self-diffusion.

**Conclusions**

In the present work, we developed a physically justified method to describe macroscopic properties of inorganic ICs in the molten state. Our approach is motivated by the significant degree of covalence in the alkali metal - halogen bonding, as suggested by the coupled cluster theory. The atomistic calculations of the macroscopic properties can be performed within our model using the standard techniques. Unlike the conventional description, our systems contain a certain percentage of dynamically assigned neutral particles. These neutral particles decrease the electrostatic energy density, hence increasing mobility of all atoms and lowering the system viscosity. The proposed method provides a reliable and well-justified atomistic-level description of ICs.

## Acknowledgments

Financial support of the NSF grant CHE-1050405 is gratefully acknowledged.

**FIGURE CAPTIONS**

**Figure 1**. Partial atomic charges on the Na and K atoms in the sodium and potassium chlorides *versus* the inter-atomic distance. The Hirshfeld charges are depicted by red circles, and the Mulliken charges are depicted by green squares. The lines are present to guide an eye.

**Figure 2**. (a) The interaction energies between cation and anion *versus* the inter-atomic distance. (b) The fraction of electrostatic interaction with respect to the total interaction energy. The results for NaCl and KCl are depicted by red circles and green squares, respectively.

**Figure 3**. Schematic of the free energy profile as a function of a reaction coordinate. The red dashed line with a single minimum corresponds to a covalent/ionic pair in the crystal (solid) state. In the molten (liquid) state, the covalent/ionic state, $|e,\Psi\rangle$, is metastable (black curve), whereas the states corresponding to neutral (left) and charged (right) atoms become stable.

**Figure 4**. Shear viscosity of the NaCl (red circles) and KCl (green squares) melts at 1080 K as a function of the fraction of neutral sodium and potassium atoms in the simulated systems.

**Figure 5**. Diffusion constants of the NaCl (red circles) and KCl (green squares) melts at 1080 K as a function of the fraction of neutral sodium and potassium atoms in the simulated systems.

FIGURE 1

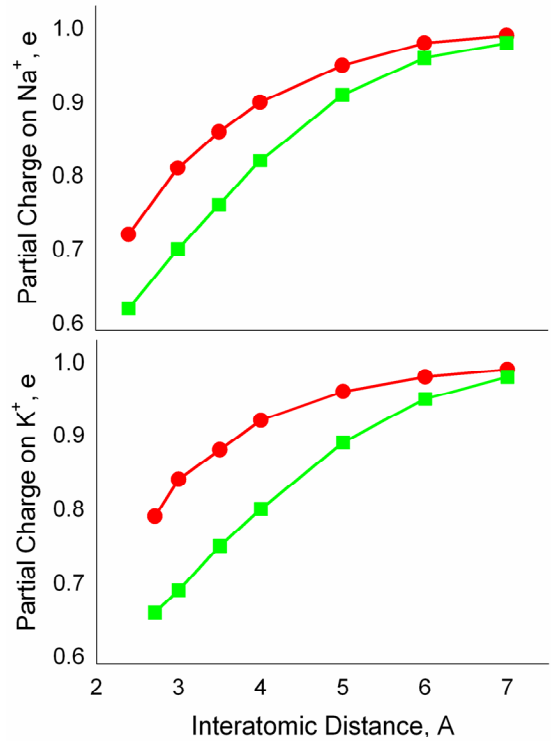

FIGURE 2

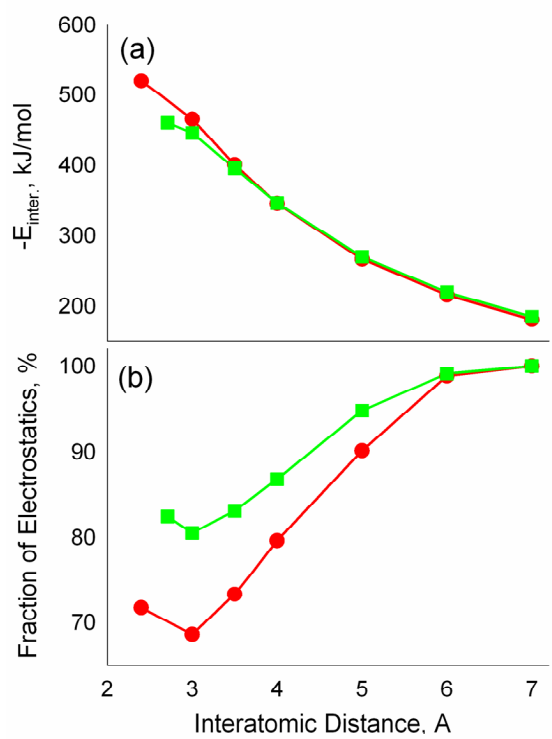

FIGURE 3

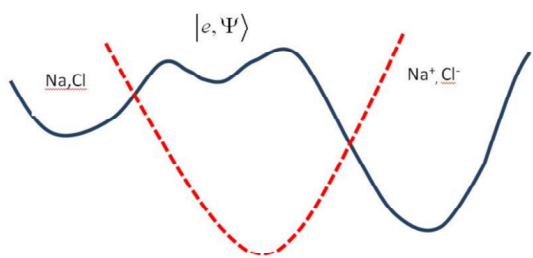

FIGURE 4

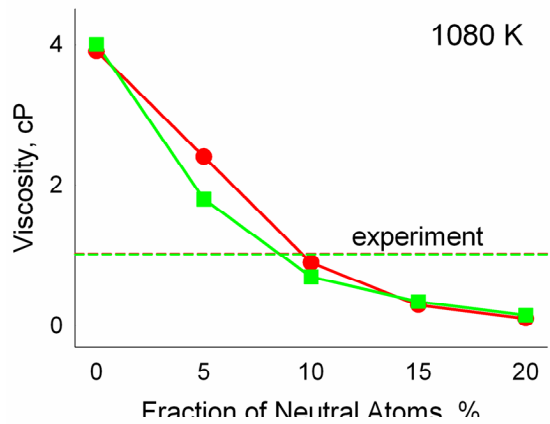

FIGURE 5

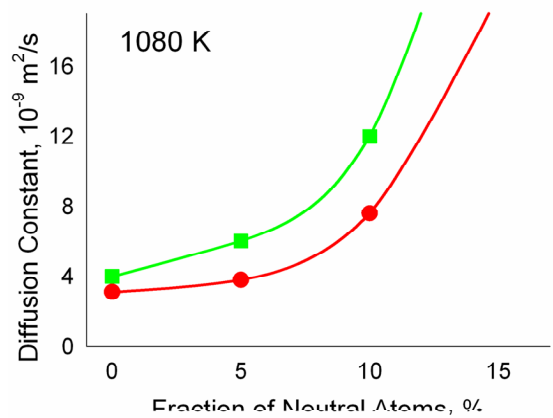